# Indirect measurement of infrared absorption spectrum through thermal emission of meta-cavity array

QIONG QIONG CHU, FENG YUAN ZHANG, YE ZHANG, SHI NING ZHU, AND HUI LIU*

*National Laboratory of Solid State Microstructures, School of Physics, Collaborative Innovation Center of Advanced Microstructures, Nanjing University, Nanjing, Jiangsu 210093, China.*

**Abstract:** Controlling thermal emission is essential for various infrared spectroscopy applications. Metasurfaces can be utilized to control multiple degrees of freedom of thermal emission, enabling the compact thermal emission materials and devices. Infrared spectroscopy such as FTIR (Fourier transform infrared spectroscopy), usually requires external infrared radiation source and complex spectroscopic devices for absorption spectrum measurement, which hinders the implementation of integrated compact and portable measurement equipment. Measuring absorption spectrum through the thermal emission of pixelated thermal emitter array can facilitate the integration and miniaturization of measurement setup, which is highly demanded for on-chip spectroscopy applications. Here, we experimentally demonstrate an integrated technology that allows for indirect measurement of the absorption spectrum through the thermal emission of meta-cavity array. This indirect measurement method opens a new avenue for compact infrared spectroscopy analysis.

## 1. Introduction

The control of thermal emission is important for the development of various infrared applications, such as thermophotovoltaic device[1-3], heat management[4-6], thermal camouflage[7, 8], radiative cooling[9-12], infrared radiation source[13, 14] and so on. Metasurfaces, as an effective way to manipulate multiple degrees of freedom of optical waves, have been widely investigated for many applications, such as meta-lens[15, 16], biosensors [17-19] and perfect absorbers[20, 21]. In recent years, metasurface based thermal emitters [22-36] have attracted tremendous attention due to their remarkable capability to tune thermal emission with multiple degrees of freedom, such as wavelengths[37], polarizations[38], radiation angles [39], bandwidth [40] and spatial radiation patterns [41]. Through metasurface array, pixelated thermal emitters[42] can further enable the integration of multifunctional infrared devices, such as spectroscopy applications.

Infrared spectroscopy [43, 44] is a common way to obtain the absorption spectrum through transmission T and reflection R spectrum measurement, as A=1-T-R, which usually requires complex spectroscopic equipment and infrared radiation source during the measuring process. According to Kirchhoff's law, the emissivity of materials can be equivalent to their absorptivity under the thermal equilibrium condition. Utilizing single thermal emitter [45, 46] as an infrared radiation source, the characteristic absorption bands of molecular can be detected through emission spectrum measurement while an infrared spectrometer or simplified spectroscopic equipment is needed. It is quite necessary to invent the integrated and compact absorption spectrum measurement technique without the need of an external infrared radiation source and infrared spectrometer. Through the thermal emission of pixelated thermal emitter array, it is possible to obtain the spectral absorption information from spatially distributed thermal emitter pixels. Indirect measurement of the infrared absorption spectrum can be realized without requirement of external infrared radiation source and infrared spectrometer. This way can further realize the miniaturization and integration of measurement device.

In this work, we demonstrate a pixelated meta-cavity array to realize a thermal emission microchip with emission wavelengths covering 7.8-12 μm infrared range. Based on this microchip design and the thermal imaging approach, we propose a highly integrated indirect infrared absorption spectrum measurement technique without the need of external infrared radiation source and infrared spectrometer. Through the thermal imaging of the microchip with or without the polymer molecule layer, the equivalent absorption spectrum of polymer molecules can be indirectly obtained from the variations of thermal emission energy. On the basis of the pixelated meta-cavity array design, demonstrated absorption spectrum measurement technique enables the integration of the infrared radiation source and the spectroscopic chip, simplifying the measurement system and reducing the operating costs. Also, in the measuring process, the direct contact between the detected sample and the chip can be avoided. Benefiting from these advantages, this novel technique can find many potential applications based on compact infrared spectroscopy, such as organic molecular sensing and environment monitoring.

## 2. Theoretical analysis

### 2.1 Thermal emission microchip design

Here, a meta-cavity array based thermal emission microchip is designed with each meta-cavity pixel possessing one emission peak. By integrating meta-cavity pixels, this microchip can simultaneously perform as an integrated infrared radiation source and infrared spectrometer, providing pixelated spectroscopic analysis. Designed microchip is composed of a nanohole metasurface array, a gold mirror, and a dielectric Si layer sandwiched between them, as shown in Fig. 1(a). Each metasurface of meta-cavity is artificially designed to tune the FP mode to obtain desired resonant wavelength. Polarized nanohole designs are chosen because their resonant wavelengths can be continuously tuned as the nanohole length increases. Through continuously modulating the nanohole parameters, a multiband microchip can be realized. Specifically, the unit cell period along x axis and the width of nanoholes are fixed as $Px=1.6$ μm, $w=0.2$ μm. The length of nanoholes depends on the unit cell period along y axis, defined as $L=Py*0.5$. We set Py as 2, 2.2, 2.4, 2.6, 3, 3.4, 3.8, 4.2 and 4.6 μm respectively, corresponding to nine different metasurfaces. The thickness of Au/Si/Au layers is set as 70 nm/0.96 μm/100 nm. Simulated absorption spectra of the meta-cavity array under x-polarized incidence are shown in Fig. 1(c). We can see that with the increase of the nanohole length, linearly tunable resonant wavelengths covering 7.8-12 μm are achieved, thus enabling the realization of multiband infrared radiation source. In contrast, under y-polarized incidence, nanohole metasurfaces cannot be excited and the corresponding absorption (gray lines) is close to zero. This polarization-selective property can be flexibly tuned by manipulating the long-axis orientation of the nanoholes.

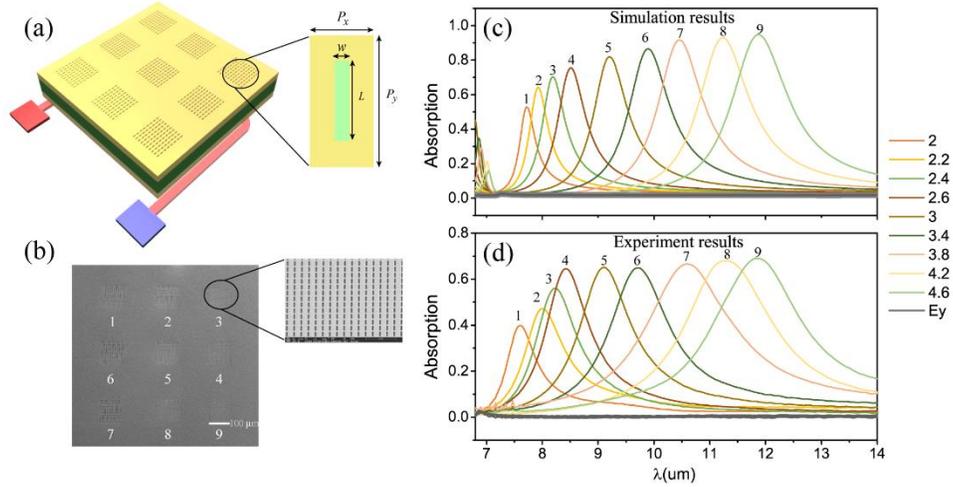

Fig. 1. Meta-cavity array based thermal emission microchip. (a) Sketch of designed thermal emission microchip. (b) Micrograph of fabricated meta-cavity array. The inset of (a) and (b) depict the unit cell of meta-cavity array and zoomed-in SEM image of fabricated meta-cavity. (c) Simulated absorption spectra of meta-cavity array under x-polarized incidence. The gray lines present the simulated absorption spectra of meta-cavity array under y-polarized incidence. (d) Measured absorption spectra of meta-cavity array.

*2.2 Equivalent absorptivity calculation*

According to Kirchhoff's law, the thermal emissivity ε of meta-cavity pixels is equal to the absorptivity A under the thermal equilibrium condition. Therefore, we can obtain the thermal emissivity and corresponding thermal emission energy of each meta-cavity pixel through its absorption spectrum. By analyzing the thermal emission intensity change before and after the thermal emission signal of meta-cavity pixels passes through the absorptive layer above it, we can obtain the equivalent absorption spectrum of the detected absorptive layer. Specifically, the variations of the thermal emission intensity of meta-cavity pixels can be represented by the variations of their thermal emission energy. The thermal emission energy of meta-cavity pixels in infinitesimal wavelength intervals can be calculated as $PB_i = \varepsilon_i M_{bb}(\lambda_i, T)(\Delta\lambda_i)$ $(i=1,2,3...)$, where ε represents the emissivity of meta-cavity pixel, $M_{bb}$ is the calculated blackbody emissivity according to Planck's law. $T$ is the temperature of the microchip and $i$ represents the number of the infinitesimal wavelength intervals. Next, when the absorptive layer is placed above the microchip, the thermal emission energy of meta-cavity pixels can be recalculated as $PB_i' = (1 - A_i - R_i)\varepsilon_i M_{bb}(\lambda_i, T)(\Delta\lambda_i)$ $(i=1,2,3...)$. Here $A_i$ and $R_i$ represent the absorptivity and reflectivity of the absorptive layer. When the reflection of absorptive layer is negligible ($R_i \approx 0$), we can obtain the equivalent absorptivity by $A_i = 1 - PB_i'/PB_i$. Through combining the equivalent absorptivity of each meta-cavity pixel, the equivalent absorption spectrum of the absorptive layer can be obtained. It should be noted that the corresponding tendency of absorption spectrum can also be obtained even if there is some reflection at the interface of absorptive layer.

## 3. Experimental results and discussions

*3.1 Infrared spectral properties of thermal emission microchip*

In order to verify the wavelength-selective property of the designed thermal emission microchip, we fabricated the pixelated metasurface array on the top gold film through focused ion beam. Specifically, the etching depth of nanoholes is set as 70 nm. More fabrication details can be found in Supplement 1. Each metasurface of meta-cavity pixel is 100 μm × 100 μm sized and the SEM (scanning electron microscope) image of fabricated sample is displayed in Fig. 1(b). Through FTIR, we measured the absorption spectra of meta-cavity array under the excitation of x-polarized incidence at room temperature, as shown in Fig. 1(d). It can be seen that the experimental results agree well with the theoretical results. With the increase of period Py and the nanohole length L, the absorption peaks of meta-cavity array gradually red shift from 7.8 to 12 μm. This multi-resonance characteristic verifies the feasibility of designed microchip as a multiband and polarized infrared radiation source. In addition, the Q factor of the absorption peaks of meta-cavity array in experiments is lower than that of simulation results because of the inevitable errors caused in the fabrication process. The difference between measured absorption spectra and simulated results of No.1 and No.4 meta-cavities is mainly caused by the broaden nanohole width in fabrication.

### 3.2 Absorption spectrum measuring of PDMS

Next, the infrared characteristics of the thermal emission microchip are investigated by utilizing a long-wave thermal camera (7.6-14 μm). The measurement system is composed of an electric heating stage and a thermal camera, as shown in Fig. 2. The stage can provide uniform heating for the microchip. An absorptive molecular layer is inserted between the microchip and the camera by a rotating robotic arm without direct contact with the microchip. The thermal emission signal from the microchip, passing through the polymer layer, is detected by the thermal camera. Then, the thermal image is obtained for analysis. Reasonably, the vibrational information of detected molecular is encoded into the thermal images, as shown in the Fig. 2. It should be noted that the microchip and the polymer layer possess independent emission properties due to different temperatures. Benefiting from the integrated infrared radiation source and spectroscopic chip, the measurement system provides new insights for the design of on-chip infrared spectroscopy devices.

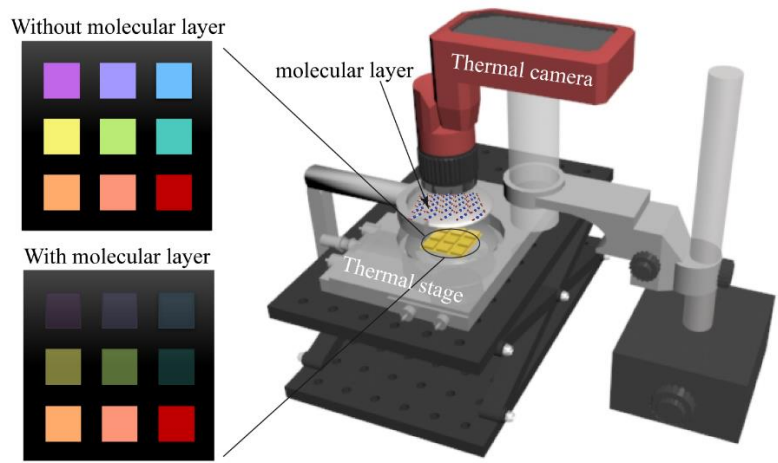

Fig. 2. Schematic diagram of the absorption spectrum measurement setup. The vibrational information of detected molecular is encoded into the thermal images taken with or without the molecular layer. The false color of the meta-cavity pixels from violet to red represents the emission wavelengths from 7.8 to 12 μm.

Firstly, we choose the PDMS (polydimethylsiloxane) organic molecular as the absorptive layer to be detected. According to the Kirchhoff's law, the emission spectrum of PDMS was measured with FTIR and utilized as the standard absorption spectrum, as shown in Fig. 3(d). According to the imaging principle of the thermal camera, the emission temperature of meta-cavity pixels in the thermal imaging is proportional to their thermal emission energy. Correspondingly, we can investigate the thermal emission energy variations of meta-cavity pixels through their emission temperature variations in thermal imaging. Then, the experimental equivalent absorptivity of the molecular layer can be calculated as $A_i = 1 - Tem_i'/Tem_i$ where the $Tem_i'$ ($Tem_i$) represents the emission temperature of meta-cavity pixels in the thermal image with (without) molecular layer. Measured reflection spectrum of PDMS layer can be found in Supplement 1. Low reflectivity of PDMS layer enables the feasibility for indirect measurement of absorption spectrum.

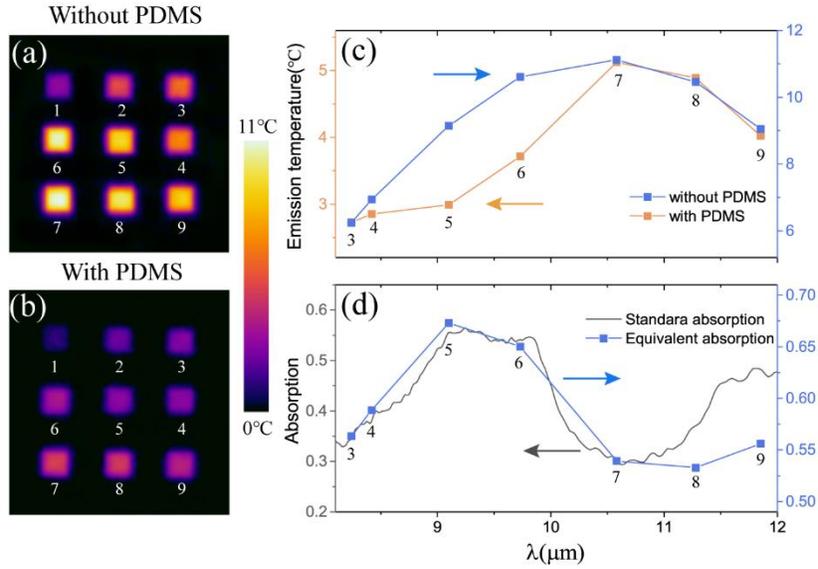

Fig. 3. Absorption spectrum measuring of PDMS. (a) The thermal image taken without molecular layer. (b) The thermal image taken with PDMS layer. (c) Measured emission temperature of meta-cavity pixels in thermal images with and without PDMS layer. (d) Standard absorption spectrum and the equivalent absorption spectrum of PDMS layer.

In the thermal imaging experiments, the temperature of microchip was raised to 100 ℃ with an electric heating stage. Then two thermal images were taken without or with the PDMS layer, as shown in Fig. 3 (a, b), respectively. For the case without PDMS layer, the meta-cavity pixels with different emission capabilities display different emission temperatures in Fig. 3(a). For the case with PDMS layer in Fig. 3(b), the emission temperature of all meta-cavity pixels in the thermal image is reduced due to the molecular absorption. In order to obtain more accurate data, the average emission temperature $Tem'$ and $Tem$ of each meta-cavity pixel in two thermal images are extracted and given in Fig. 3(c). The area for averaging emission temperature is the area of fabricated nanohole metasurface of each meta-cavity. For clearly comparing the thermal emission difference of each meta-cavity pixel in the imaging, we set the average emission temperature of each meta-cavity pixel and its corresponding central emission wavelength as the y and x axis, respectively. Here, it should be noted that the emission temperature shown in the figure is the result after subtracting the background emission temperature (the emission temperature of metal surface around nanohole metasurfaces). Comparing the average emission temperature results in Fig. 3(a) and Fig. 3(b), it is clear that the emission temperature of No.5-6 meta-cavity pixels undergoes a significant decrease after the polymer layer is added. This means that the central emission wavelengths of No. 5-6 meta-cavity pixels equal to the absorption wavelengths of PDMS molecular. Comparing the two thermal images allows us to make an intuitive and quick measurement of the absorption spectrum of PDMS molecular. According to the derived emission temperature data of different meta-cavity pixels, we can obtain the experimental equivalent absorption spectrum of PDMS molecular without FTIR, as shown in Fig. 3(d). It can be seen that the equivalent absorption spectrum of PDMS is consistent with the standard absorption spectrum of PDMS measured with FTIR. Therefore, the absorption spectrum of PDMS molecular can be simply obtained from thermal imaging. In Fig. 3(d), the deviation between the experimental equivalent absorption spectrum and the standard absorption spectrum of PDMS mainly comes from the limited measured data due to the insufficient number of meta-cavity pixels. In the future, the measurement accuracy can be further improved through increasing the number of meta-cavity pixels. Furthermore, the absorptivity discrepancies between these two curves are mainly caused by the different detectors and data processing programs. The reflection of the emitted signal from meta-cavity pixels at the molecular layer has very slight impact on the absorption spectrum.

*3.3 Absorption spectrum measuring of ETFE, PTFE and PVDC*

In order to demonstrate the universality of proposed imaging-based absorption spectrum measurement technique, we also detected three other types of organic molecular, ETFE (ethylene-tetra-fluoro-ethylene), PTFE (Poly-tetra-fluoroethylene) and PVDC (Polyvinylidene chloride), as shown in Fig.4. Measured reflection spectra of those molecular layers can be found in Supplement 1. In Fig. 4(d), the emission temperature of No.1-7 meta-cavity pixels undergoes a significant decrease due to the absorption of molecular layer. Comparing with the standard absorption spectrum, the equivalent absorption spectrum of ETFE clearly shows the absorption bands in the same wavelength range (7.6-10.5 μm). In other two experiments of PTFE and PVDC, the emission temperature of No.2-5 meta-cavity pixels and No.5-6 meta-cavity pixels respectively undergo a clear decrease after adding the molecular layer, as shown in Fig. 4(e, f). The equivalent absorption spectra of PTFE and PVDC show the absorption bands of 7.6-9 μm and 9.2-9.8 μm respectively. These results are consistent with the absorption spectra measured with FTIR, verifying the feasibility and universality of this measurement method. More details can be found in Supplement 1. In our experiments, this simple repeatable measurement process can further improve the utilization and practicality of proposed absorption spectrum measurement technique.

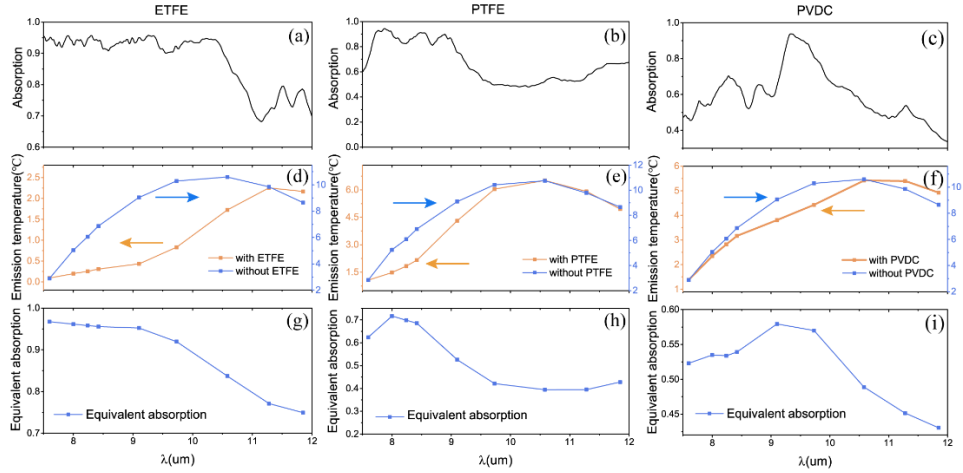

Fig. 4. Absorption spectrum measuring of ETFE, PTFE and PVDC. Absorption spectrum of molecular layer measured by FTIR (a-c); Measured emission temperature of meta-cavity pixels in thermal images with and without molecular layer (d-f); Equivalent absorption spectrum of molecular layer obtained from thermal imaging method (g-i).

It should be noted that the detection wavelength resolution is limited by the linewidth of the emission peaks of meta-cavity pixels due to metal ohmic loss in the long-wave infrared domain. For example, in Fig. 4(a), the vibrational modes of ETFE molecules with narrow linewidth near 11.5 μm cannot be detected in the experiment. In future, the wavelength resolution can be improved through integrating high-Q designs, such as a BIC array based on dielectric metasuface pixels. Based on above detection results, we can conclude that this imaging-based absorption spectrum measurement technique can efficiently and rapidly obtain the equivalent absorption spectrum of various organic molecular. In addition, thanks to the greatly simplified measurement system, this integrated technique can be flexibly applied to many other compact infrared applications, such as integrated micro-molecular sensing and material analysis.

## 4.  Conclusion

In summary, we proposed an integrated and compact technique to realize indirect infrared absorption spectrum measurement based on a pixelated meta-cavity array. By utilizing long-wave thermal camera, we can analyze the thermal emission of each meta-cavity pixel through thermal imaging with or without the absorptive layer. Then, the equivalent absorption spectrum of analyte compounds can be obtained when the reflection is negligible. In experiments, we have realized the absorption spectrum measurement of four kinds of molecules and the results are consistent with the absorption spectra obtained through FTIR. The designed thermal emission microchip can perform as an infrared radiation source and a spectroscopic chip simultaneously, realizing integrated compact on-chip absorption spectrum measurements. Although the wavelength resolution of this absorption spectrum measurement technique is limited due to ohmic loss in metal, future designs based on dielectric metasurfaces with less loss can be expected to improve it. This proposed imaging-based absorption spectrum measurement technique provides a potential technique for the miniaturization and integration of infrared applications, such as on-chip micro-molecular sensing, environmental monitoring and medical diagnosis.

**Funding.** This work was financially supported by the National Natural Science Foundation of China (Nos. 92163216, 92150302 and 62288101).

**Disclosures.** The authors declare no conflicts of interest.

**Data availability.** Data underlying the results presented in this paper are not publicly available at this time but may be obtained from the authors upon reasonable request..

**Supplemental document.** See Supplement 1 for supporting content.

# Indirect measurement of infrared absorption spectrum through thermal emission of meta-cavity array

QIONG QIONG CHU, FENG YUAN ZHANG, YE ZHANG, SHI NING ZHU, AND HUI LIU[*]

*National Laboratory of Solid State Microstructures, School of Physics, Collaborative Innovation Center of Advanced Microstructures, Nanjing University, Nanjing, Jiangsu 210093, China.*
*liuhui@nju.edu.cn*

## Absorption spectrum measuring of PDMS, ETFE, PTFE and PVDC

1. **Thermal emission microchip fabrication**

Firstly, through magnetron sputtering (ULVAC CS-200z), three layers of Au (70 nm) /Si(0.96 μm) /Au(100 nm) are sequentially deposited on the Si substrate. Then 3×3 nanohole metasurfaces is etched from the top gold film by a focused ion beam (FIB dual-beam FEI Helios 600i, 30 keV, 100 pA). In fabrication process, the etching length of nanohole L are varied from 1 μm to 2.3 μm while the etching depth is set as 70 nm. Each metasurface of meta-cavity pixel is 100 μm×100 μm sized.

2. **Molecular layer preparation**

PTFE, ETFE and PVDC molecular layers are commercial processed thin films with thickness of 30μm, 25μm, and 10μm respectively. PDMS molecular layer with thickness of 2μm (diluted PDMS solution) is spin coated on the KBr layer at 1200 rpm for 2 mins. The KBr layer with thickness of 2 mm is transparent to infrared light within 7.6-14 μm range.

3. **Optical characterization**

The reflection spectra (R) of the meta-cavity pixels under x-polarized incidence were firstly measured using the FTIR spectrometer. The reflection signals were collected using a Hyperion 2000 IR microscope with a liquid-nitrogen-cooled HgCdTe (MCT) detector. Measured reflection spectra were normalized with respect to a gold mirror. The absorption spectrum (A) of each meta-cavity pixel is derived on the basis of the measured reflection spectrum (R) by A = 1 − R. The standard absorption spectra of molecular layers were obtained by measuring their thermal emission spectra. After heating to 100℃, the emission signal from molecular sample was collected through the heater window and then analyzed by FTIR spectrometer. Measured emission spectra were normalized with respect to a carbon power sample (can be approximately seem as blackbody).

4. **Reflection of molecular layers**

It can be seen that these molecular layers have a relatively low reflectivity compared to their absorption components. Also, it should be noted that the oscillating resonant peaks in reflection spectrum are mostly caused by the interference effect in the thin film, which is hard to see in the thermal detecting experiments due to the commonly incoherent properties of thermal emission. Therefore, the reflection at the interface of molecular layer in the thermal detecting experiments has slighter influence on the absorption spectrum. This negligible reflection enables the indirect absorption spectrum measurement of molecular layer.

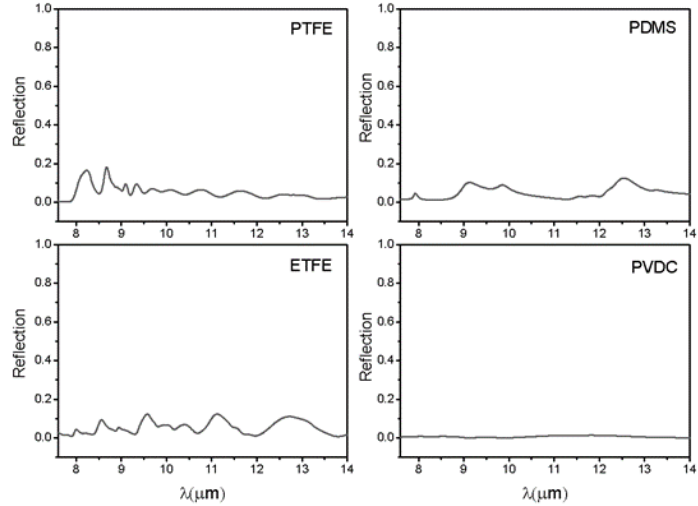

Fig. S1. Reflection spectrum of PTFE, PDMS, ETFE and PVDC.

## 5. Calculated thermal emission energy and measured emission temperature of meta-cavity pixels

To visualize the proportional relationship[1, 2] between the thermal emission energy and the emission temperature of meta-cavity pixels, we calculated the thermal emission energy of each meta-cavity pixel with and without the polymer molecular layer. Calculated thermal emission energy, measured emission temperature in thermal images and the corresponding thermal images are shown in Fig. S2, S3, S4 and S5.

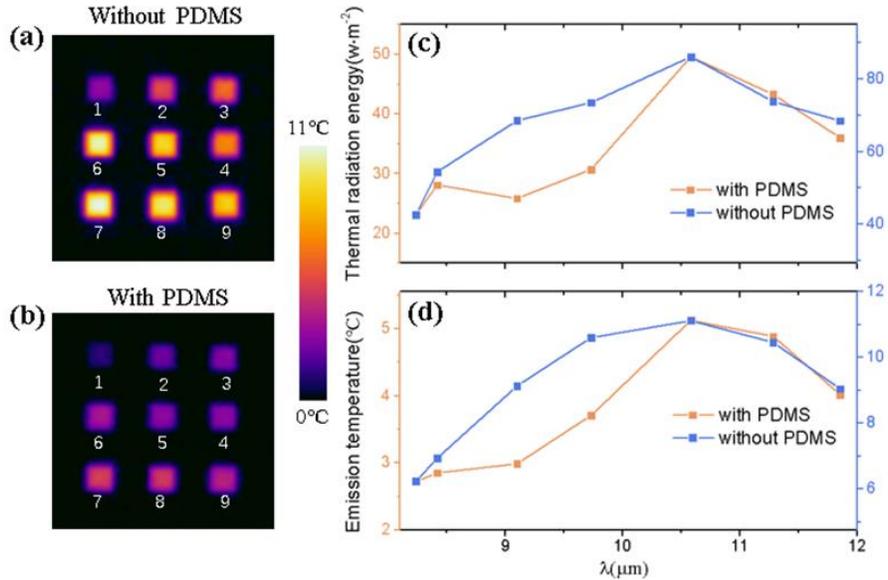

Fig. S2. Absorption spectrum measuring of PDMS. (a) The thermal image taken without molecular layer. (b) The thermal image taken with PDMS layer. (c) Calculated thermal emission energy of meta-cavity

pixels with and without PDMS layer. (d) Measured emission temperature of meta-cavity pixels in thermal images with and without PDMS layer.

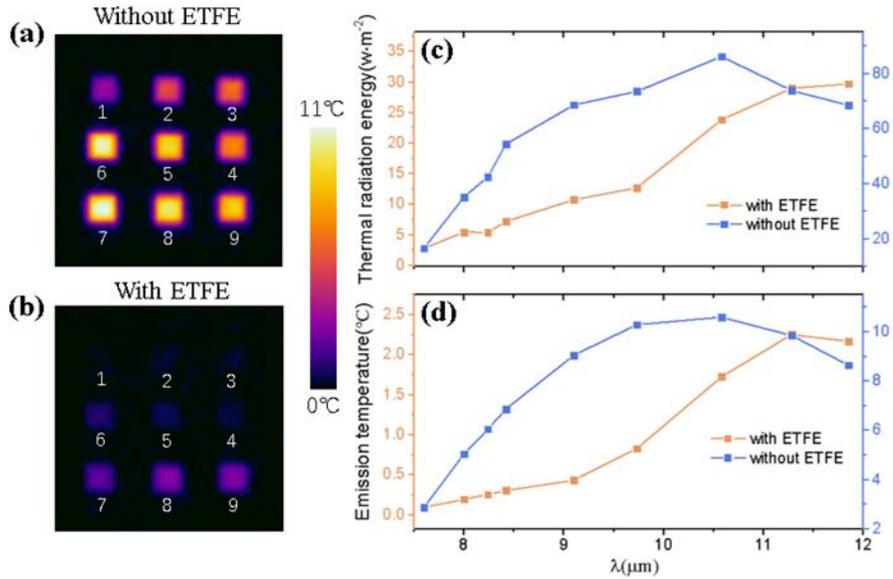

Fig. S3. Absorption spectrum measuring of ETFE. (a) The thermal image taken without molecular layer. (b) The thermal image taken with ETFE layer. (c) Calculated thermal emission energy of meta-cavity pixels with and without ETFE layer. (d) Measured emission temperature of meta-cavity pixels in thermal images with and without ETFE layer.

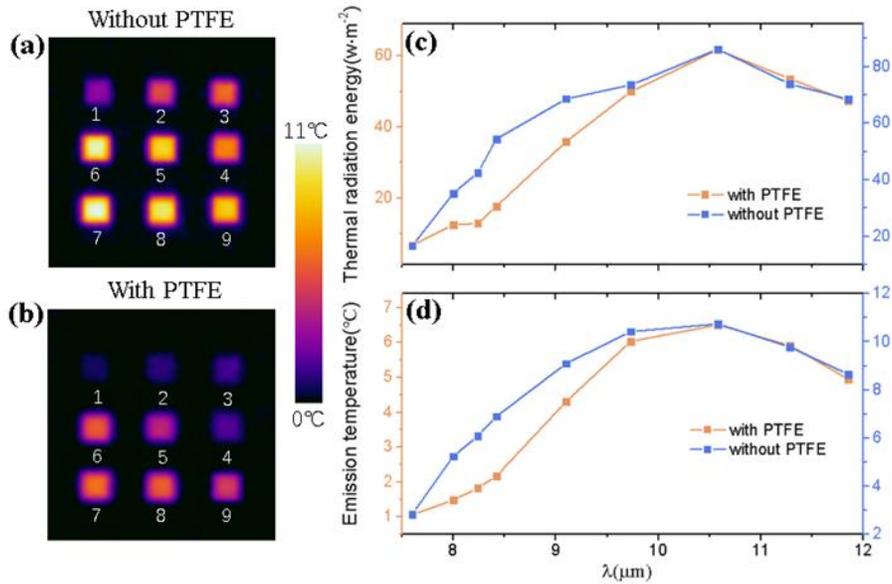

Fig. S4. Absorption spectrum measuring of PTFE. (a) The thermal image taken without molecular layer. (b) The thermal image taken with PTFE layer. (c) Calculated thermal emission energy of meta-cavity pixels with and without PTFE layer. (d) Measured emission temperature of meta-cavity pixels in thermal images with and without PTFE layer.

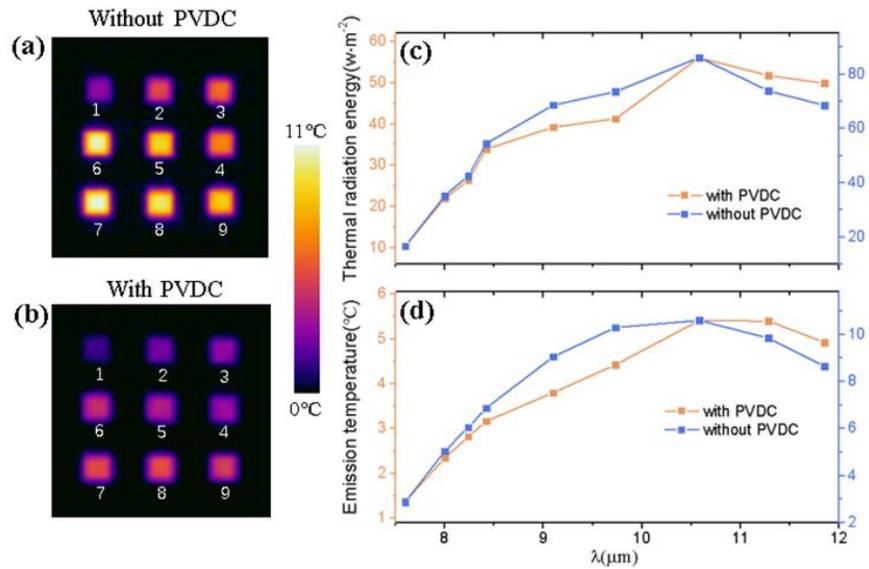

Fig. S5. Absorption spectrum measuring of PVDC. (a) The thermal image taken without molecular layer. (b) The thermal image taken with PVDC layer. (c) Calculated thermal emission energy of meta-cavity pixels with and without PVDC layer. (d) Measured emission temperature of meta-cavity pixels in thermal images with and without PVDC layer.